# Spin Hall magnetoresistance in Pt/(Ga,Mn)N devices

J. Aaron Mendoza-Rodarte[1,2]*, Katarzyna Gas[3,4], Manuel Herrera-Zaldívar[2], Detlef Hommel[5], Maciej Sawicki[3,6], and Marcos H. D. Guimarães[1]*

[1]*Zernike Institute for Advanced Materials, University of Groningen, 9747 AG Groningen, The Netherlands*

[2]*Centro de Nanociencias y Nanotecnología-Universidad Nacional Autónoma de México, Ensenada, 22800-Baja California, México*

[3]*Institute of Physics, Polish Academy of Sciences, Aleja Lotnikow 32/46, PL-02668 Warsaw, Poland*

[4]*Center for Science and Innovation in Spintronics, Tohoku University, Katahira 2-1-1, Aoba-ku, Sendai 980-8577, Japan*

[5]*Lukasiewicz Research Network - PORT Polish Center for Technology Development, Stabłowicka 147, Wrocław, Poland*

[6]*Research Institute of Electrical Communication, Tohoku University, Katahira 2-1-1, Aoba-ku, Sendai 980-8577, Japan*

**\*Corresponding authors:**

J. Aaron Mendoza-Rodarte, j.a.mendoza.rodarte@rug.nl; Marcos H. D. Guimarães, m.h.guimaraes@rug.nl

Diluted magnetic semiconductors (DMS) have attracted significant attention for their potential in spintronic applications. Particularly, magnetically-doped GaN is highly attractive due to its high relevance for the CMOS industry and the possibility of developing advanced spintronic devices which are fully compatible with the current industrial procedures. Despite this interest, there remains a need to investigate the spintronic parameters that characterize interfaces within these systems. Here, we perform spin Hall magnetoresistance (SMR) measurements to evaluate the spin transfer at a Pt/(Ga,Mn)N interface. We determine the transparency of the interface through the estimation of the real part of the spin mixing conductance finding $G_r = 2.6 \cdot 10^{14}$ $\Omega^{-1}$ m$^{-2}$, comparable to state-of-the-art yttrium iron garnet (YIG)/Pt interfaces. Moreover, the magnetic ordering probed by SMR above the (Ga,Mn)N Curie temperature $T_C$ provides a broader temperature range for the efficient generation and detection of spin currents, relaxing the conditions for this material to be applied in spintronic devices.

Recent advances in the manipulation of electron charge and spin have highlighted spintronics as a key area of research[1,2]. This interest stems from the potential of spintronic devices to address the limitations of traditional charged-based devices, such as high energy consumption[3]. Among various materials, diluted magnetic semiconductors (DMS), those systems where a fraction of cations are substituted by magnetic elements, stand out as promising candidates for spintronic applications due to their unique integration of semiconductor and magnetic properties[4]. Particularly, Mn-doped compounds have received significant attention from the research community since the formation of a ferromagnetic order at technologically relevant temperatures was predicted[5] and realized[6]. Among these compounds, single-phase Mn-doped GaN epitaxial layers stand out for their insulating and short-range ferromagnetic character at low temperatures. It presents a rich playground to explore magnetic interactions of magnetic ions in a semiconductor lattice. A brief discussion on the ferromagnetic mechanisms in (Ga,Mn)N is included in the supplementary material. The combination of these factors, coupled with GaN well-established roles in optoelectronics[7], high-frequency[8,9], and power electronics[10] could provide substantial technological benefits, particularly in the development of GaN-based spintronic devices[11]. In fact apart from exploitation of the magnetoelectric effect in (Ga,Mn)N so far [12], the study of GaN-based DMS has been focused on optimizing growth conditions[13] and magnetic properties[14,15], while spin transport dynamics received comparatively less attention[16–20]. It is therefore timely and important to exploit spintronic and magnonic[21] properties of this insulating ferromagnetic material in an aim to foster the development of all-nitride low power information processing and dissipationless communication means based on spin waves propagation.

Of particular importance for nonvolatile magnetic data storage applications, the field of spin-orbit torques (SOTs) focuses on manipulating the magnetization in thin film heterostructures via electric currents, providing a promising way to manipulate magnetization at the nanoscale[22]. This technology relies on transferring spin angular momentum from a normal metal into a ferromagnet, using spin-charge interconversion effects to exert a torque on the magnetization[22]. Significant advancements in SOT have catalyzed the development of non-volatile memory devices[23,24]. However, these devices depends on the efficient generation, transport, and detection of spin currents. While material research with high spin-orbit coupling[25] or orbital angular momentum effects[26] addresses spin current generation, transport issues are managed by identifying materials with high interface spin transparency, indicated by substantial spin mixing conductance values ($G_{\uparrow\downarrow}$). Spin Hall magnetoresistance (SMR) has emerged as a key technique for assessing spin transport properties at heterostructure interfaces through electrical methods[27]. Despite these advances, there is still a significant gap in understanding spin transport properties in heterostructures that utilize transition metal-doped GaN, which is crucial for SOT applications.

In this study, we investigate spin Hall magnetoresistance (SMR) within a Pt/(Ga,Mn)N heterostructure. By performing angle-dependent magnetoresistance measurements, we determine the spin mixing conductance ($G_{\uparrow\downarrow}$), a critical parameter that dictates the transport of spin information through the interface between the two materials. Additionally, we provide another assessment of the Curie temperature of the magnetic layer solely through electrical means, via temperature-dependent measurements. Our results provide valuable insights for the development of devices that incorporate GaN-based DMS for advanced spintronic applications.

To characterize the Pt/(Ga,Mn)N interface via SMR, we fabricated a 6 nm-thick Pt Hall bar on a 100 nm-thick single-phase epitaxy $Ga_{0.922}Mn_{0.078}N$ film. The film was grown using plasma-assisted molecular beam epitaxy (MBE) on a 3 μm-thick GaN(0001) template, which was deposited on *c*-oriented 2-inch sapphire substrates. Detailed methodologies for the growth and magnetic and crystallographic characterization of these films are discussed in detail in reference [14]. The Hall bar, measuring 25 μm in wide and 200 μm in length, was fabricated by conventional lithography processes. Prior to the deposition of Pt through $Ar^+$ plasma d.c. sputtering, the (Ga,Mn)N surface was cleaned using an $Ar^+$ mild etching process, detailed in supplementary material. Electrical measurements were performed by rotating an external magnetic field within the plane of the device. We utilized conventional lock-in techniques for the measurements, with a bias current ($I_{bias}$) of less than 0.7 mA at a frequency of 187.77 Hz. This setup allowed us to measure the transverse resistance ($R_{xy}$), which is perpendicular to the current path. This layout of the electrical interconnects is illustrated in Fig. 1(a).

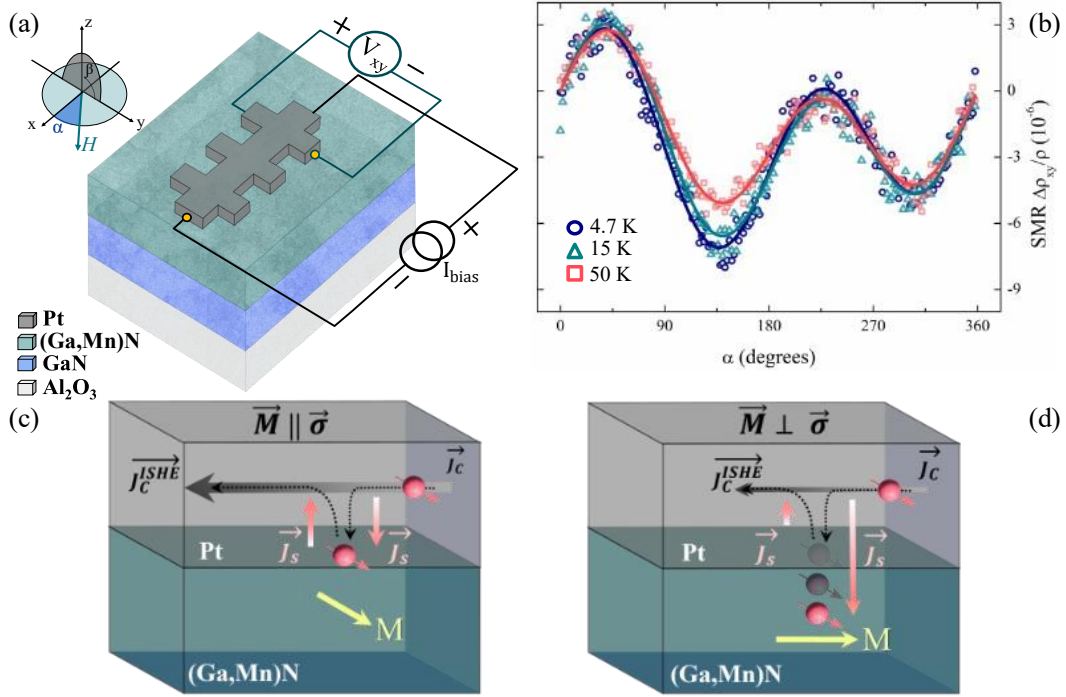

FIG. 1. *(a) Schematics of the devices and electrical measurement configuration, illustrating the polarity of the contacts and the definition of the rotating external magnetic field angle α. (b) Angle dependence of the SMR signal in the transverse geometry as a function of α. The measurements were conducted in the temperature range of 4.7 to 50 K using an external magnetic field of 600 mT. The bulk resistivity of Pt has a magnitude of $1.58 \cdot 10^{-7}\ \Omega\ m$. A baseline has been removed so that the relative changes in resistivity are zero at α=0 degrees. (c-d) Schematic representation of SMR. (c) Low-resistance configuration where $\vec{M}||\vec{\sigma}$. (d) High-resistance configuration where $\vec{M} \perp \vec{\sigma}$.*

The SMR measurements offer insights into the magnetic ordering at the surface of the (Ga,Mn)N thin film at low temperatures. Fig. 1(b) illustrates the dependence of the magnetic field angle (α) on the relative changes in the transverse resistivity, $\Delta\rho_{xy}/\rho$. Here, $\Delta\rho_{xy}/\rho$ is defined as $(R_{xy} - R_{0,xy})/(R_{0,xx}/4.8)$ where $R_{xy}$ is the transverse resistance, and $R_{0,xy}$ and $R_{0,xx}$ are the transverse and longitudinal resistances at zero magnetic field with values between 126 to 127 and 1.2 to 1.3 Ω, respectively, and 4.8 is the geometrical factor of the Hall bar (length/width). The observed changes in resistivity can be explained as follows: when a charge current ($J_c$) flows through Pt, the spin Hall effect (SHE) induces a transverse spin current ($J_s$) towards the (Ga,Mn)N/Pt interface. At this interface, spin accumulation interacts with the local magnetic moments of (Ga,Mn)N. If the spin polarization $\vec{\sigma}$ is parallel to the magnetization direction $\vec{M}$ of the ferromagnet ($\vec{M}||\vec{\sigma}$), the electrons' angular momentum is reflected, resulting in a low-resistance configuration due to the inverse spin Hall effect (ISHE) [see Fig. 1(c)]. Conversely, when $\vec{M}$ is perpendicular to $\vec{\sigma}$ ($\vec{M} \perp \vec{\sigma}$) [refer to Fig. 1(d)], the electrons' angular momentum is absorbed, leading to a high-resistance configuration. In transverse SMR measurements, the interaction of the generated spin current with the (Ga,Mn)N magnetization $\vec{M}$ at the interface results in a reorientation of the outgoing spin current polarization $\vec{\sigma}$. At angles α = 45 and 135°, this leads to positive and negative transverse voltage, respectively[28]. Our experimental observations align with this theory, showing maximum resistance at α = 45° and minimum at α = 135°, similar to findings in YIG/Pt systems[29]. While anisotropic magnetoresistance (AMR) could theoretically contribute to the observed magnetoresistance signal[30], this is highly improbable in our system. The significant resistivity mismatch between the Pt layer and the (Ga,Mn)N layer makes it less likely that AMR at the interface is the dominant mechanism. Our experimental results, which show a clear dependence of the transverse resistance on the angle α in line with SMR predictions, further support this conclusion. The dependence of the resistivity in the transverse geometry ($\rho_T$) with respect to the magnetization components along the different directions is given by[27]:

$$\rho_T = \Delta\rho_1 m_x m_y + \Delta\rho_2 m_z + \Delta\rho_{Hall} B_z, \qquad (1)$$

where $\Delta\rho_1$ and $\Delta\rho_2$ represents relative changes in resistivity and $m_x$, $m_y$, and $m_z$ are the components (unit vectors) of magnetization in the $\hat{x}$-, $\hat{y}$- and $\hat{z}$-direction, respectively. The angular dependencies are defined by the components $m_x$, $m_y$, and $m_z$, where $m_x = \cos(\alpha)\cos(\beta)$, $m_y = \sin(\alpha)\cos(\beta)$, and $m_z = \sin(\beta)$, with $\beta$ representing the azimuthal angle – i.e. the angle with respect to the out-of-plane (OOP) direction. Using these definitions, we can express the change in the transverse resistivity as follows[27]:

$$\Delta\rho_{xy}/\rho = \Delta\rho_{xy,1}\frac{1}{2}\sin(2\alpha) - \Delta\rho_{xy,2}\sin(\beta), \qquad (2)$$

where $\Delta\rho_{xy,1}$ and $\Delta\rho_{xy,2}$ represent the amplitudes of the relative change in resistivity. This equation effectively describes our observed signals and fits our measurements well. Additionally, the term $\Delta\rho_{Hall} B_z$ accounts for the ordinary Hall effect occurring in Pt when a magnetic field is applied in the $\hat{z}$-direction. The $\sin(\beta)$ component observed in our results likely results from OOP component due to misalignment of the Hall bar with respect to the in-plane field. At 4.7 K, the values for the amplitudes $\Delta\rho_{xy,1}$ and $\Delta\rho_{xy,2}$ are 7.2·10⁻⁶ and 1.8·10⁻⁶, respectively, indicating a clear dominance of the in-plane component.

Using SMR measurements, we investigate the ferromagnetic-paramagnetic transition of (Ga,Mn)N solely through electrical methods. To this end, we perform a temperature-dependent measurement ranging from 4.7 to 290 K. Fig. 2 displays the behavior of magnitudes $\Delta\rho_{xy,1}$ and $\Delta\rho_{xy,2}$ derived from Eq. (1) across different temperatures. As the temperature increased from 4.7 to 10 K, $\Delta\rho_{xy,1}$ fluctuated between $7.2\cdot10^{-6}$ and $7.3\cdot10^{-6}$ with the maximum value at 10 K. Above 10 K, a decrease in $\Delta\rho_{xy,1}$ was noted, dropping to $6.81\cdot10^{-6}$ at 15 K. This reduction in the SMR signal suggests a $T_C$ between 10 to 15 K, aligning well with the $T_C$ of $13.0 \pm 0.3$ K established by SQUID magnetometry[14]. The signal diminished above 50 K, and data fitting became unreliable due to a significant decrease in the signal-to-noise ratio (SNR). Nevertheless, it is important to note that substantial SMR signals have been observed from 15 to 50 K, a range dominated by the paramagnetic phase. This magnetic ordering is similar to that seen in other Curie-like paramagnetic insulators such as $Gd_3Ga_5O_{12}$ (GGG), $CoCr_2O_4$, and recently $NdGaO_3$[31–35]. In the specific case of Pt/GGG interface, the SMR arises from the interaction between the conduction-electron spins in Pt and the paramagnetic spins $S$ within GGG via the interface exchange interaction, which applies torque on $S$. Regarding the fitted OOP component, its temperature dependence displayed a monotonic behavior, with $\Delta\rho_{xy,2}$ values between 1.6 and $1.9 \cdot 10^{-6}$. This points towards an ordinary Hall contribution in our transverse measurements, which can be attributed to a sample misalignment.

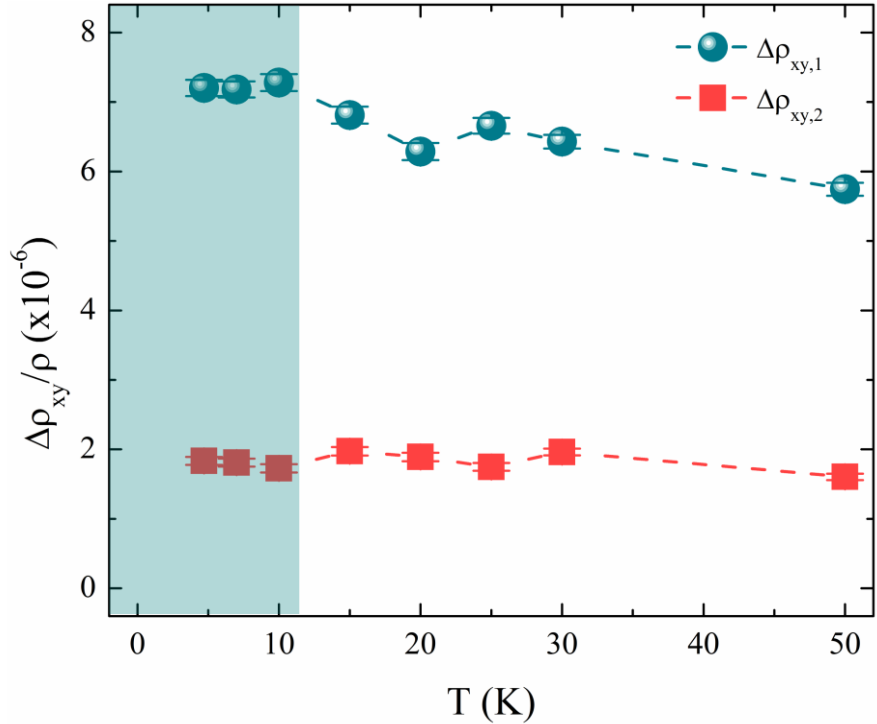

FIG. 2. *Temperature dependence of the relative changes of the resistivity of $\Delta\rho_{xy,1}$ and $\Delta\rho_{xy,2}$. The green shadow highlights the decrease of the SMR signal observed above 10 K. All the amplitudes were extracted from Eq. (1) using measurements performed at 600 mT.*

The magnetic field dependence of $\Delta\rho_{xy,1}$ and $\Delta\rho_{xy,2}$ demonstrates quadratic and linear behaviors, respectively, as shown in Fig. 3a and 3b. We can attibute the former dependency to the percolating character of ferromagnetism in our material, indicating that only at $T = 0$ all the spins present in the sample are ferromagnetically coupled. On increasing $T$ some of the spins start decoupling such that at its equilibrium $T_C$ only about 20% of spins remain in the infinite cluster[36]. The remaining spins reside in *finite* ferromagnetic clusters of different sizes. These magnetic granules are characterized by their own magnetic moments and are responsible for the glassy characteristics like blocking (i.e. as in the material being far from thermal equilibrium). Such material can therefore mimic a ferromagnet well above its equilibrium $T_C$ when it is probed on sufficiently short time scales, as during SMR experiments. The relatively small Magnetoresitance (MR) amplitude could be also attributed to Hanle Magnetoresistance (HMR), which has been reported at Pt/YIG[37]. However, HMR typically persists even at room temperature, above the Curie temperature of (Ga,Mn)N, as it primarily depends on spin accumulation in the Pt layer. In contrast, our experiments show that the MR effect decreases significantly above the Curie temperature, which supports SMR as the dominant mechanism rather than HMR. Additionally, at 600 mT we do not observe a saturation state, a finding corroborated by SQUID magnetometry *M-H* curves, which indicate that even at $\mu_0 H = 7$ T at $T = 2$ K the system still exhibits a positive slope (as indicated in the supplementary material Fig. S2). We also note that we expect that the MR amplitude will increase by one or two orders of magnitude near the magnetization saturation point of (Ga,Mn)N ($\mu_0 H = 7$ T). The coercive field ($H_c$) values from SQUID magnetometry for similar samples suggest an increase of the average magnetization above 200 mT, aligning with the clear SMR signal observed at 360 mT in our results (Fig. 3a). Additionally, the linear dependence seen in $\Delta\rho_{xy,2}$ aligns with the expected behavior of the ordinary Hall component. To confirm that our measurements are within the linear regime, we conducted bias current ($I_{bias}$) dependence tests. Fig. 3c shows the transverse voltage ($V_{xy}$) as a function of α, and Fig. 3d displays the $I_{bias}$ dependency of the relative change in transverse voltage $\Delta V_{xy}$, where a data point represents the amplitude of the angle-dependent voltage, obtained by a fit to $V_{xy} = \Delta V_{xy,1}\frac{1}{2}\sin(2\alpha) - \Delta V_{xy,2}\sin(\beta)$. Linear fits applied to both the in-plane ($\Delta V_{xy,1}$) and out-of-plane ($\Delta V_{xy,2}$) components confirm the linear behavior, with no evident heating effects.

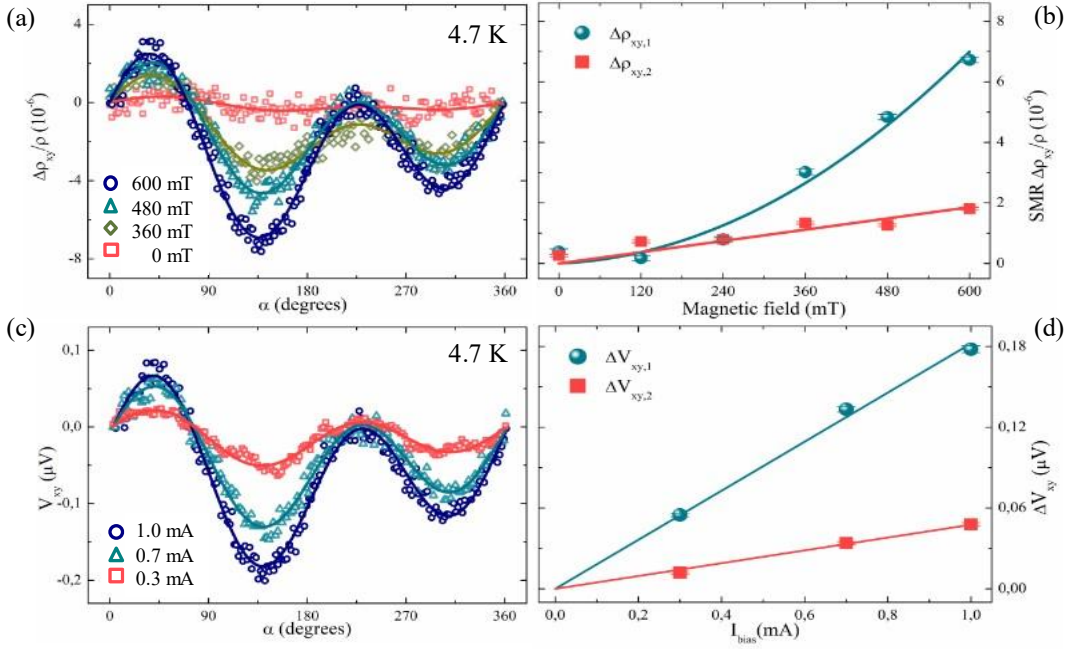

FIG. 3. *(a) Relative change of the transverse resistivity $\rho_{xy}$ as a function of angle α at different values of magnetic field B and for bias current $I_{bias}$=1.0 mA. (b) Magnetic field dependence of $\Delta\rho_{xy,1}$ and $\Delta\rho_{xy,2}$. (c) Angle dependence of $V_{xy}$ at different $I_{bias}$. (d) Current dependence of $\Delta V_{xy}$. All the measurements were performed at a fixed temperature of 4.7 K. A baseline has been removed so that the relative changes in resistivity are zero at α=0 degrees.*

The transfer of spin angular momentum at the interface of the heterostructure is quantified by the spin mixing conductance, $G_{\uparrow\downarrow}$, which we estimate from our experimental results. From Eq. (1), we can express $\Delta\rho_1$ and $\Delta\rho_2$ as follows[27]:

$$\frac{\Delta\rho_1}{\rho} = \theta_{SH}^2 \frac{\lambda}{d_N} \mathrm{Re}\left(\frac{2\lambda G_{\uparrow\downarrow} \tanh^2\frac{d_N}{\lambda}}{\sigma + 2\lambda G_{\uparrow\downarrow} \coth\frac{d_N}{\lambda}}\right), \qquad (3)$$

$$\frac{\Delta\rho_2}{\rho} = -\theta_{SH}^2 \frac{\lambda}{d_N} \mathrm{Im}\left(\frac{2\lambda G_{\uparrow\downarrow} \tanh^2\frac{d_N}{\lambda}}{\sigma + 2\lambda G_{\uparrow\downarrow} \coth\frac{d_N}{\lambda}}\right). \qquad (4)$$

Here, $\theta_{SH}$ represents the spin Hall angle, λ the spin relaxation length, $d_N$ the thickness, and σ the conductivity of Pt. $G_{\uparrow\downarrow}$ denotes the relaxation of the spin current polarization component transverse to the magnetization at the Pt/(Ga,Mn)N interface[38] and is defined as the sum of its real and imaginary parts ($G_{\uparrow\downarrow} = G_r + iG_i$). In our calculations, we have neglected the imaginary part ($G_i$) based on the assumption that its contribution is significantly smaller than that of the real part ($G_r$). This assumption is supported by several reports, which suggest that in similar insulating systems, the real part dominates the total spin mixing conductance[38–40].

The Pt conductivity σ is calculated from the full resistivity tensor, yielding a value of 6.28·10[6]

$\Omega^{-1}$ m$^{-1}$, consistent with literature values [41–43]. Assuming constant values of $\theta_{SH} = 0.08$, $\lambda = 1.1 \pm 0.3$ nm [39,44], taking $d_N = 6$ nm, and using our SMR data at 4.7 K and B = 600 mT, we calculate the real part of the spin mixing conductance, which under $G_{\uparrow\downarrow} \approx G_r$ condition attains a value of $2.6 \cdot 10^{14}$ $\Omega^{-1}$ m$^{-2}$. This value is comparable to those reported for other heavy metal/magnetic insulator interfaces such as YIG/Pt[39].

Our study shows that the Pt/(Ga,Mn)N interface exhibits spintronic properties comparable to state-of-the-art systems, such as YIG/Pt interfaces[44], as evidenced by similar spin mixing conductance values. This indicates the strong potential of magnetically-doped GaN for spintronic and magnonic applications such as magnonic transistors[45] and SOT devices[46]. Furthermore, the magnetic ordering probed by SMR in (Ga,Mn)N above its equilibrium Curie temperature extends sizably the temperature range for efficient generating and detecting spin currents in this material. This broadens the operational parameters for its application in spintronic devices with higher operating temperatures. Our findings contribute significantly to understanding spin transport in (Ga,Mn)N-based devices and set the stage for exploring various device architectures for spintronic applications based on the nitride family.

## SUPPLEMENTARY MATERIAL

See supplementary material for a brief discussion of the ferromagnetic mechanism in (Ga,Mn)N insulating systems, a detailed description of Pt/(Ga,Mn)N device fabrication, and SQUID magnetometry measurements, which includes Ref.[12,14,36,47–65].

## ACKNOWLEDGMENTS

We acknowledge M. Cosset-Chéneau and J. J. L. van Rijn for helpful discussions and critically reading the manuscript, and J. Holstein, H. de Vries, F.H. van der Velde, H. Adema, and A. Joshua for technical support. This work was supported by the Dutch Research Council (NWO – OCENW.XL21.XL21.058), the Zernike Institute for Advanced Materials, the research program "Materials for the Quantum Age" (QuMat, registration number 024.005.006), which is part of the Gravitation program financed by the Dutch Ministry of Education, Culture and Science (OCW), the National Science Centre (Poland) through project OPUS (DEC-2018/31/B/ST3/03438), and the European Union (ERC, 2D-OPTOSPIN, 101076932). Views and opinions expressed are however those of the author(s) only and do not necessarily reflect those of the European Union or the European Research Council. Neither the European Union nor the granting authority can be held responsible for them. JAMR is grateful to CONAHCYT for a graduate research fellowship (no. CVU 655591). The device fabrication and characterization were performed using Zernike NanoLabNL facilities.

# AUTHOR DECLARATIONS

### Conflict of Interest

The authors have no conflicts to disclose.

### Author Contributions

**J. Aaron Mendoza-Rodarte:** Conceptualization (equal); Methodology (lead); Validation (lead); Formal analysis (lead); Investigation (lead); Data Curation (lead); Writing – Original Draft (lead); Visualization (equal). **Katarzyna Gas:** Resources (supporting); Writing – Review & Editing (supporting). **Manuel Herrera-Zaldívar:** Supervision (supporting); Writing – Review & Editing (supporting). **Detlef Hommel:** Resources (supporting); Writing – Review & Editing (supporting). **Maciej Sawicki:** Resources (supporting); Writing – Review & Editing (supporting). **Marcos H. D. Guimarães:** Supervision (lead); Project administration (lead); Conceptualization (equal); Visualization (equal); Resources (lead); Writing – Review & Editing (lead); Funding acquisition (lead).

# DATA AVAILABILITY

The data that support the findings of this study are available from the corresponding author upon reasonable request.